\documentclass{revtex4-2}

\usepackage{graphicx}
\usepackage{dcolumn}
\usepackage{bm}
\usepackage{hyperref}
\usepackage[mathlines]{lineno}
\usepackage[dvipsnames]{xcolor}
\usepackage{color}
\usepackage{amsfonts}
\usepackage{amsopn}
\usepackage{amsmath}

\begin{document}

\preprint{APS/123-QED}

\title{Stability of Orbital Angular Momentum Modes in Conventional and Ring-Core Optical Fibers}

\author{
Hassan Asgharzadeh B.$^{1,*}$}
\email{hassan.asgharzadehbastehdimi@tuni.fi}
\author{Regina Gumenyuk$^{1}$}
\author{Marco Ornigotti$^{1}$}

\date{\today}

\affiliation{$^{1}$Tampere University, Photonics Laboratory, Physics Unit,
Tampere FI-33720, Finland}

\begin{abstract}

Modal dynamics in multimode optical fibers are fundamentally governed by the interplay between phase matching and intermodal coupling. In this work, we investigate the fundamental coupling mechanisms in ring-core fibers and identify a geometry-induced coupling suppression that significantly reduces coupling coefficients between orbital angular momentum modes compared to conventional multimode step-index fibers. This mechanism provides a clear physical explanation for the superior transmission stability of orbital angular momentum modes in ring-core fibers and establishes specific geometric constraints for the design of ring-core fiber-based mode converters.
\end{abstract}

\maketitle
\section{Introduction}

The exponential growth in global data traffic has driven optical communications toward mode-division multiplexing using Orbital Angular Momentum (OAM) modes, due to the theoretically infinite alphabet they are capable to provide \cite{huang2015mode}. However, the practical deployment of OAM-based systems, whether for long-distance transmission or signal processing, is fundamentally governed by the stability of these modes as they propagate through fibers \cite{7523310}. Under realistic operating conditions, optical fibers are always subject to structural perturbations such as micro-bending, core offsets, and elliptical deformations. These imperfections break the cylindrical symmetry of the waveguide, inducing modal interactions \cite{snyderLove,marcuse2013theory,rao2024high} that degrade the modal purity of the propagating OAM beams.

To address these challenges, Ring-Core Fibers (RCFs) have emerged as the preferred fiber geometry for OAM beam transmission than conventional multimode Step-Index fibers (SIFs)\cite{alexeyev1998optical,ma2021generation}. Furthermore, twisting the fiber structure can be beneficial for improving both transmission and mode conversion. In transmission scenarios, for example, a small twist can reduce the degeneracy between eigenmodes \cite{davtyan2020robust,ye2016excitation} and, additionally, the spinning nature of the fiber averages linear birefringence and other imperfections \cite{barlow1981birefringence}, thus stabilizing the propagation of OAM modes. In contrast, for mode converting scenarios, fibers are twisted around phase matching rates to form Chiral Long Period Fiber Gratings (CLPFGs), aiming at converting an initially excited mode into higher-order OAM modes \cite{shvets2009polarization,huang2021excitation,wu2022low,wang2023high}. In general, both scenarios are governed by the same physical parameters, namely the phase matching condition and the modal coupling coefficient between the interacting modes. 

The stability of OAM modes in RCFs has generally been attributed to the large effective index separation ($\Delta n_{\mathrm{eff}}$) between different modes, which effectively breaks the phase-matching condition \cite{6732926,bozinovic2013terabit,ma2021generation}, or to the possibility of achieving better contrast between core and cladding, than in standard fibers \cite{bozinovic2013terabit}. While $\Delta n_{\mathrm{eff}}$ is critical, the effect of the modal coupling coefficients (which is determined by the overlap of the modal fields with the perturbations) can also play an important role. In a conventional solid-core fiber, these interactions usually occur at a single boundary, i.e., the core-cladding interface. In an RCF, on the other hand, the optical field interacts with perturbations at both boundaries between the core and the inner and the outer cladding. To the best of our knowledge, the impact of this Dual-Boundary Effect (DBE) on the net coupling strength has not been investigated before, and remains therefore unexplored. 

In this paper, we present a unified analysis of modal coupling in twisted RCFs, identifying a fundamental DBE that affects OAM mode interactions. Using the Coupled-Mode Equations (CMEs) model we recently derived for twisted RCFs \cite{asgharzadeh2026optical}, we calculate modal couplings arising from various perturbation sources such as core offset and non-uniformity (e.g., tapering), and we demonstrate that the refractive index changes at the inner and outer boundaries of an RCF inherently have the opposite impact on the overlap integral required to calculate modal coupling coefficients. This results in a geometric mitigation mechanism that can significantly affect the net coupling coefficient. We compare these results with conventional SIFs to show that RCFs are intrinsically more resilient to symmetric perturbations, thus showing how this geometric filtering effect provides a physical basis for the superior stability of RCFs for transmission applications. In addition to that, our calculations also show how DBE can be used as a guideline for RCFs engineering for mode conversion applications.

Our work is organized as follows: in Sect.\ref{section2} we briefly review the relevant theoretical framework of coupled mode equations in the presence of twisting and tapering, that constitute the foundational frameowrk of our analysis. In Section \ref{section3} we discuss the dual-boundary mitigation mechanisms and we compare its effect between RCFs and conventional fibers, providing numerical simulations to corroborate our findings. Finally, Conclusions are drawn in Sect. \ref{section4}.

\section{Theoretical Framework} \label{section2}

\subsection{Coupled Mode Theory}

The dynamics of OAM modes in twisted fibers can be analyzed using the following generalized set of CMEs \cite{asgharzadeh2026optical}

 \begin{equation}
     \frac{d{\widetilde{a}}_j(z)}{dz}=\sum_{q}{{\widetilde{a}}_q\left(z\right)}\left\{i\mathbb{K}_{jq}e^{i\left(\beta_q-\beta_j\pm\left\{J_q-J_j\right\}\Upsilon\right)z}+\Omega_{jq}e^{i\left(\beta_q-\beta_j\right)z}\right\},
     \label{eq:cmes}
 \end{equation}
where $\mathbb{K}_{jq}$ represents twist-induced coupling coefficient between modes $j$ and $q$ propagating through the perturbed (e.g., twisted core offset) fibers, $\Upsilon$ is the twist rate (defined as $\Upsilon=2 \pi/\Lambda$, with $\Lambda$ being the twist pitch), $\Omega_{jq}$ is the modal interaction due to any perturbation (such as tapering or micro-bends) along the propagation direction $z$, $\beta$ is the propagation constant of untwisted modes in the reference fiber, $J_j$ and $J_q$ are the total angular momenta of interacting modes $j$ and $q$, respectively, and $\widetilde{a}(z)\in\mathbb{C}$ is the local mode amplitude propagating in the perturbed fiber. Notice, that by setting $\Upsilon=0$, the analysis is reduces to untwisted fibers.

As detailed in \cite{asgharzadeh2026optical}, the explicit expression for $\mathbb{K}_{jq}$ can be expressed as
\begin{equation}
\mathbb{K}_{jq}=\frac{k_0}{4} \left(\frac{\varepsilon_0}{\mu_0}\right)^{1/2}\iint \boldsymbol{\mathbb{\xi}}_j\cdot\left[\boldsymbol{\delta\varepsilon}\right]\cdot\boldsymbol{\mathbb{\xi}}_q^{*}r dr d\varphi, 
\label{eq:ccoetw}
\end{equation}
where $\boldsymbol{\mathbb{\xi}}_j$ and $\boldsymbol{\mathbb{\xi}}_q$ are the  normalized 3D electric field distributions of the reference modes $j$ and $q$, respectively, $k_0$, is the free-space wavenumber, and $\left[\boldsymbol{\delta\varepsilon}\right]$ is the perturbation tensor.
The explicit expression for $\Omega_{jq}$ is instead given by \cite{asgharzadeh2026optical, snyderLove}
\begin{equation}
\Omega_{jq}=\frac{1}{4} \frac{k_0}{\beta_{j}-\beta_{q}}\left(\frac{\varepsilon_0}{\mu_0}\right)^{1/2} \iint \boldsymbol{\mathbb{\xi}}_{jt}\cdot\boldsymbol{\mathbb{\xi}}_{qt}^{*} \frac{\partial n^2}{\partial z} r dr d\varphi,
\label{eq:ccoeta}
\end{equation}
with the subscript $t$ denoting the transverse part of the local electric field $\boldsymbol{\mathbb{\xi}}$. It is also important to notice, that for longitudinal, $z$-dependent perturbations, such as tapering or micro bends, the refractive index of the fiber becomes $z$-dependent \cite{snyderLove}. 
\subsection{DBE in RCFs}
In this section, we rigorously analyze the coupling coefficients induced by perturbations in both conventional SIFs and RCFs. Specifically, we investigate a geometry-driven mechanism intrinsic to RCFs that mitigates modal coupling strength. We demonstrate that this suppression arises from DBE. While perturbations in SIFs are localized to a single core-cladding interface, RCFs involve perturbation contributions at both the inner and outer radii of the ring-guide, leading to a modification of the net coupling coefficient.
\begin{figure}[!t]
\centering\includegraphics[width=12cm]{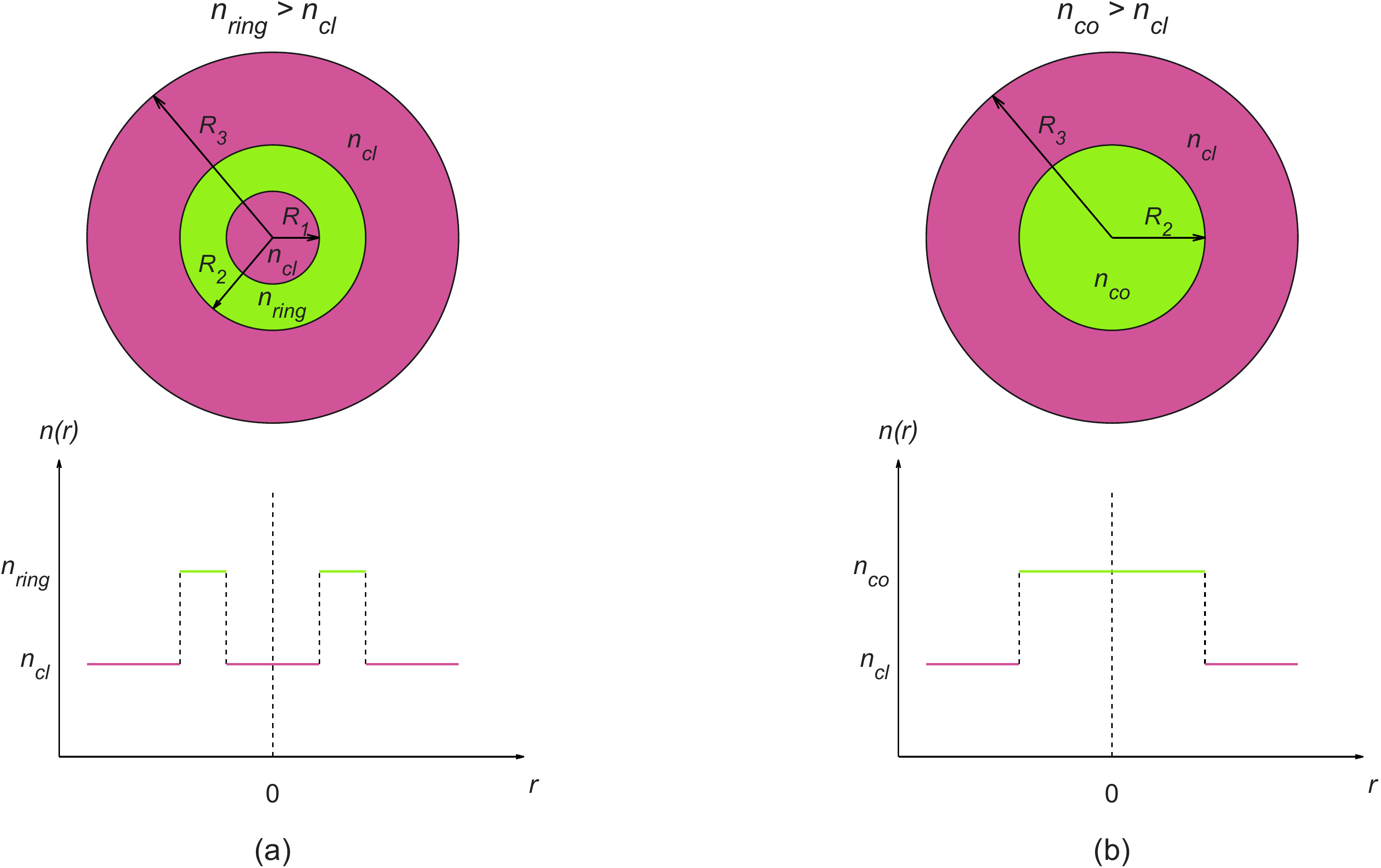}
\caption{Cross-sectional views and refractive index distributions of the optical fibers analyzed in this study: (a) multimode RCF and (b) multimode SIF. For consistency, the reference SIF is modeled as the limiting case where $R_1 \to 0$ ($\rho \to 0$), while keeping the outer radius $R_2$ and the refractive indices unchanged. Here, $R_1$ and $R_2$ are the inner and outer radii of the ring in the RCF, $R_2$ is the core radius of the SIF, and $R_3$ is the cladding radius for both geometries. The refractive indices of the SIF core, RCF ring, and cladding are represented by $n_{\text{co}}$, $n_{\text{ring}}$, and $n_{\text{cl}}$, respectively.}
\end{figure}
We investigate modal coupling arising from two primary perturbations, including tapering and core offset. By evaluating these perturbations in both RCFs and conventional SIFs, we quantify and compare the resultant modal interactions.
\subsubsection{Taper-induced perturbation}
We begin with modal interactions arising from taper-induced perturbation. For an RCF subject to longitudinal variation, the gradient of the squared refractive index profile $\partial{n}^2/\partial z$ can be expressed as \cite{asgharzadeh2026optical}
\begin{equation}
    \frac{\partial{n}^2}{\partial z}=(n^{2}_{ring}-n^{2}_{cl})\left[\delta \left(r-R_{2}(z)\right)\frac{d R_{2}(z)}{d z}-\delta (r-R_{1}(z))\frac{d R_{1}(z)}{d z}\right],
    \label{eq:derring}
\end{equation}
where $R_{1}(z)$ and $R_{2}(z)$ are the local inner and outer radii of the ring-guide, and $n_{ring}$ and $n_{cl}$ represent the corresponding refractive indices of the ring and cladding regions, respectively (see figure~1a). For conventional SIFs, on the other hand, the gradient term simplifies to the single interface contribution as \cite{shi2016theoretical}
\begin{equation}
    \frac{\partial n^2}{\partial z}=(n^{2}_{co}-n^{2}_{cl})\delta (r-R_{1}(z))\frac{d R_{1}(z)}{d z},
    \label{eq:dercon}
\end{equation}
where $R_{1}(z)$ corresponds to the local radius of the core, and $n_{co}$ and $n_{cl}$ are the refractive indices of the core and cladding, respectively (see figure~1b). 
Substituting Eq.~(\ref{eq:derring}) and the transverse part of the normalized modal fields into Eq.~(\ref{eq:ccoeta}) yields the following analytical form of the taper-induced coupling coefficient $\Omega_{jq}$ \cite{asgharzadeh2026optical}
\begin{equation}
\begin{split}
    \Omega_{jq} &= \frac{1}{4} \frac{k_0}{\beta_{j}-\beta_{q}}\left(\frac{\varepsilon_0}{\mu_0}\right)^{1/2}(n^{2}_{\text{ring}}-n^{2}_{\text{cl}}) \biggl[ R_2(z)\frac{dR_2}{dz}F_{j,m}(R_2(z))F_{q,\nu}(R_2(z)) \\
    &\quad - R_1(z)\frac{dR_1}{dz}F_{j,m}(R_1(z))F_{q,\nu}(R_1(z)) \biggr],
\end{split}
\label{eq:ccoetaanlring}
\end{equation}
where $F_{j,m}(r)$ and $F_{q,\nu}(r)$ describe the radial field distributions for modes $j$ and $q$, characterized by radial orders $m$ and $\nu$, respectively. By applying an analogous procedure and substituting Eq.~(\ref{eq:dercon}) into Eq.~(\ref{eq:ccoeta}), the corresponding coupling coefficient for conventional SIFs instead reads
\begin{equation}
\begin{split}
    \Omega_{jq} &= \frac{1}{4} \frac{k_0}{\beta_{j}-\beta_{q}}\left(\frac{\varepsilon_0}{\mu_0}\right)^{1/2}(n^{2}_{\text{co}}-n^{2}_{\text{cl}}) \biggl[ R_1(z)\frac{dR_1}{dz}F_{j,m}(R_1(z))F_{q,\nu}(R_1(z))  \biggr].
\end{split}
\label{eq:ccoetaanlcon}
\end{equation}
A direct comparison between Eqs.~(\ref{eq:ccoetaanlring}) and ~(\ref{eq:ccoetaanlcon}) explicitly reveals DBE in RCFs, arising from the superposition of perturbations at both the inner and outer radii of the ring-guide.

\subsubsection{Core offset and ellipticity}
For conventional SIFs, the permittivity perturbation distribution, $\delta \varepsilon(r,\varphi)$, arising from a transverse core displacement can be expressed as \cite{shvets2009polarization}:
\begin{equation}
    \delta \varepsilon(r,\varphi)=(n^{2}_{\text{co}}-n^{2}_{\text{cl}})\delta (r-R_{1}) d_{N}\cos{(N \varphi)},
    \label{eq:pertoffellcon}
\end{equation}
where $N$ is the azimuthal harmonic order ($N = 1$ for core offset and $N = 2$ for core ellipticity) of the perturbation, and $d_{N}$ represents the corresponding perturbation amplitude. For RCFs, Eq.~(\ref{eq:pertoffellcon}) can be generalized as
\begin{equation}
    \delta \varepsilon(r,\varphi)=(n^{2}_{\text{co}}-n^{2}_{\text{cl}}) d_{N}\cos{(N \varphi)}\biggl[\delta (r-R_{2})-\delta (r-R_{1})\biggl].
    \label{eq:pertoffellring}
\end{equation}

By substituting Eqs.~(\ref{eq:pertoffellcon}) and (\ref{eq:pertoffellring}) into Eq.~(\ref{eq:ccoetw}) we can then evalute the induced coupling coefficients in SIFs and RCFs, respectively. Unlike the case of tapering, the full analytical derivation for these perturbations results in an algebraically extensive expressions, which we do not report here, for the sake of simplicity. Instead, we directly compare the perturbation definitions
%
in Eq.~(\ref{eq:pertoffellcon}) and Eq.~(\ref{eq:pertoffellring}), and
we quantify the impact of the DBE mechanism through the numerical evaluation of these coupling magnitudes.
\section{Numerical results and discussion}\label{section3}
\subsection{Fiber models}
To investigate the proposed dual-boundary mitigation mechanism, we establish a numerical model including two RCFs (RCF1 and RCF2) and a conventional reference SIF. The refractive index profile of the RCFs consists of a high-index guiding region ($n_{\text{ring}}$) embedded in a uniform background, where the central and outer cladding indices are matched ($ n_{\text{cl}}$).
The geometric parameters of both RCFs are standardized with a fixed outer guide radius $R_{2} = 6.0\,\mu\text{m}$ and a cladding radius $R_3 = 62.5\,\mu\text{m}$ \cite{wu2022mode,brunet2014vector}. These fibers are distinguished by their inner-to-outer radius ratio, $\rho = R_{1}/R_{2}$. Specifically, RCF1 is characterised by $\rho=0.25$, while RCF2 by $\rho = 0.50$. To ensure a rigorous analysis, the reference SIF is constructed as the limiting case where $\rho \to 0$ ($R_1 = 0$), while strictly preserving the outer radius $R_2$ and the refractive indices locked at the same values for the RCFs. This design constraint guarantees that all fibers possess an identical guide-region numerical aperture, effectively isolating the influence of the inner ring interface on the magnitude of modal couplings. 
\begin{figure}[!t]
\centering\includegraphics[width=11.5cm]{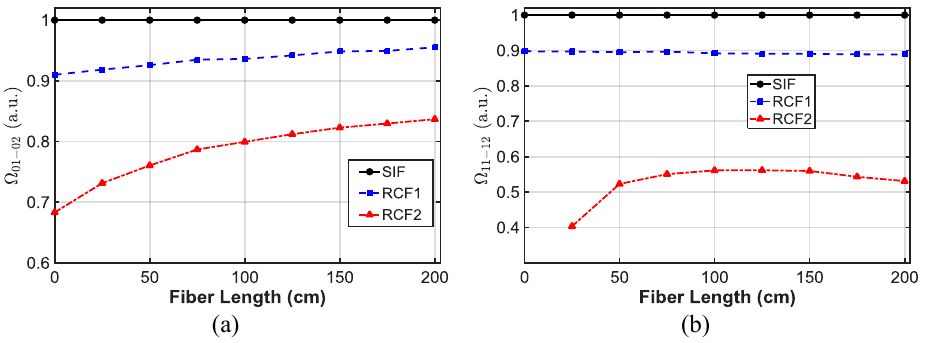}
\caption{Normalized taper-induced coupling coefficients between (a) $\text{OAM}_{01}$ and $\text{OAM}_{02}$ ($\Omega_{01-02}$), and (b) $\text{OAM}_{11}$ and $\text{OAM}_{12}$ ($\Omega_{11-12}$). The shape factor of the taper profile for both SIF and RCF geometries is set to $b_{\text{f}} = 1.5$, with a numerical aperture of $\text{NA} \approx 0.29$ and a total fiber length of $L = 2\text{ m}$.}
\end{figure}
The cross-sectional geometries of the RCFs and the reference SIF are illustrated in Figure~1(a) and figure~1(b), respectively. All calculations are performed in the continuous-wave (CW) regime at an operating wavelength of $\lambda = 1550\text{ nm}$. Without loss of generality, all OAM modes are assumed to be left-circularly polarized.

To model the tapering perturbation, we characterize the longitudinal evolution of the fiber radii, $R(z)$, using a quadratic profile defined as
\begin{equation}
R(z) = \frac{b_0 - b_f}{2L} z^2 + \frac{b_f}{2} z + R^s,
\label{eq:taperprof}
\end{equation}
where $R^s$ and $R^w$ are the radii at the narrow and wide ends of the taper, respectively, and $L$ represents the total taper length. The parameter $b_0 = 2(R^w - R^s)/L$, represents the average taper angle. The profile curvature is modulated by the dimensionless shape factor $b_f$.
\subsection{Comparative results}
\subsubsection{Taper-induced perturbation}
According to~\cite{asgharzadeh2026optical}, taper-induced crosstalk is restricted to intra-family modes differing only in radial order. We calculate coupling coefficients $\Omega\,_{01\,-\,02}$ and $\Omega\,_{11\,-\,12}$, which describe the energy transfer between guided modes $\text{OAM}_{01}$ and $\text{OAM}_{02}$, and $\text{OAM}_{11}$ and $\text{OAM}_{12}$, respectively. This condition holds for both the RCF and the conventional SIF.

The calculated taper-induced coupling coefficients are presented in figures~2(a) and (b), assuming a shape factor of $b_f = 1.5$ and a numerical aperture of $\text{NA}\approx0.29$ \cite{brunet2014vector}. The length of the tapered fiber is 2 m and $R^w = 1.25\,R^s$. Simulation results reveal that the coupling coefficients in RCF1 are larger than those in RCF2. This confirms that as the inner radius vanishes, $R_{1}\to\,0$, the modal coupling behavior in ring-core fibers smoothly approaches that of the conventional SIF. In addition, a direct comparison between figures~2(a) and (b) demonstrates that the DBE has a more pronounced impact on $\Omega\,_{11\,-\,12}$ than on $\Omega\,_{01\,-\,02}$. In RCF2, $\Omega_{11-12}$ is approximately 40--60\% smaller than that of the reference SIF (figure~2(b)), while $\Omega_{01-02}$ shows a 20--30\% reduction (figure~2(a)). This suppression serves as direct numerical evidence of the DBE. Moreover, because the taper introduces a non-uniform (longitudinally varying) perturbation, the transverse modal field distributions evolve continuously along the propagation direction, making the spatial overlap between interacting modes, Eq.~(\ref{eq:ccoeta}), $z$-dependent. As a result, the coupling suppression percentage also varies along the propagation distance $z$.

\subsubsection{Core offset}
To assess the modal coupling strength induced by transverse core displacement, we numerically evaluate the overlap integral in Eq.~(\ref{eq:ccoetw}) using Eqs.~(\ref{eq:pertoffellcon}) and~(\ref{eq:pertoffellring}). The calculated coupling coefficients $\mathbb{K}_{01-11}$, $\mathbb{K}_{11-21}$ and $\mathbb{K}_{21-31}$ are presented in figures~3(a)-(c) for a fixed offset amplitude of $d_1 = 1.0\,\mu\text{m}$.

In agreement with the case of the taper-iduced perturbation, RCF2 yields lower offset-induced coupling magnitudes than RCF1. However, we observe that the mitigation efficiency of DBE is strongly dependent on the azimuthal mode order. For interactions involving higher azimuthal orders, the suppression effect decreases. For example, figure~3(c) reveals that for $\mathbb{K}_{21-31}$, the coupling strength in RCF1 is nearly identical to that of the reference SIF ($\approx 98\%$), while RCF2 shows a reduction to $\approx 78\%$.
\begin{figure}[!t]
\centering\includegraphics[width=11.5cm]{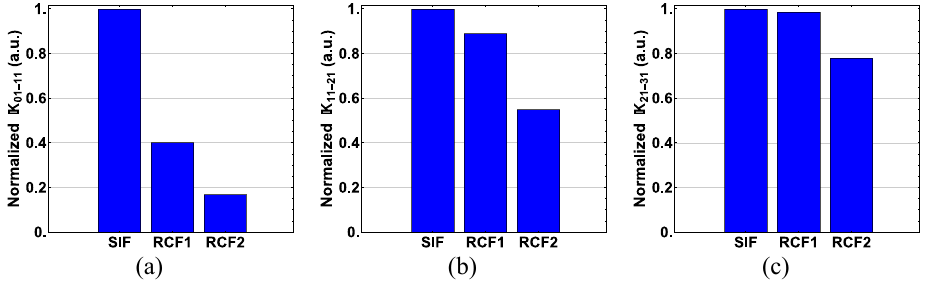}
\caption{Calculated coupling coefficients arising from a transverse core offset of $d_1 = 1.0\,\mu\text{m}$: (a) $\mathbb{K}_{01-11}$, (b) $\mathbb{K}_{11-21}$ and (c) $\mathbb{K}_{21-31}$. The numerical aperture of the guiding regions is $\approx 0.29$. As in figure~2, the conventional SIF serves as the reference one, and the coupling coefficients in RCF1 and RCF2 are compared with the corresponding ones in the reference fiber.}
\end{figure}
In contrast, for lower-order interactions such as $\mathbb{K}_{01-11}$, figure~3(a), DBE is highly effective, reducing the coupling in RCF1 and RCF2 to $\approx 40\%$ and $18\%$ of the reference one, respectively. To explain the physical origin of this mode-dependent sensitivity, we analyze the transverse intensity distributions of the supported OAM modes, $\text{OAM}_{01}$, $\text{OAM}_{11}$, $\text{OAM}_{21}$, and $\text{OAM}_{31}$, in RCF1 as illustrated in figures~4(a)–(d), respectively. 
\begin{figure}[!t]
\centering\includegraphics[width=6.5cm]{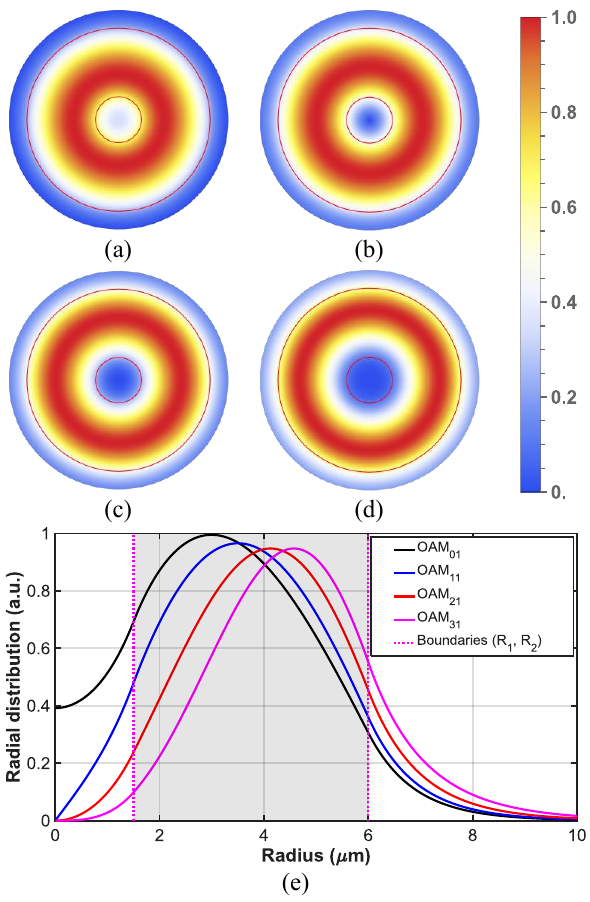}
\caption{Intensity distribution of the first four families of OAM modes supported by RCF1. (a) $\text{OAM}_{01}$, (b) $\text{OAM}_{11}$, (c) $\text{OAM}_{21}$ and (d) $\text{OAM}_{31}$. The corresponding radial distributions of these modes are shown in (e).}
\end{figure}
Inspection of figures~4(a)–(d) reveals that the radial position of the intensity maximum is strongly dependent on the topological charge. Higher OAM charge, in fact, drives the optical power away from the central axis, so that the maximum of the intensity distribution shifts further from the fiber axis, resulting in a substantial reduction of intensity at the inner radius of the guiding-ring. As a result, higher-order OAM modes exhibit reduced spatial overlap with the inner boundary interface, minimizing the contribution of the inner-radius perturbation to the total coupling. To quantify this behavior, figure~4(e) presents the normalized radial distribution of the modes shown in figures~4(a)–(d). The dashed vertical lines indicate the inner and outer radii ($R_1 = 1.5\,\mu\text{m}$ and $R_2 = 6\,\mu\text{m}$) of the ring guide in RCF1. Modal profiles explicitly confirm mode-dependent radial shift, demonstrating that the magnitude of the optical field at the inner boundary vanishes rapidly with increasing azimuthal order, resulting in the inner ring interface being less effective for higher-order modes. As illustrated in figure~4(e), the peak intensities of the four OAM modes are located at radial positions of $r_{01}^{max}\approx3\,\mu\text{m}$, $r_{11}^{max}\approx3.6\,\mu\text{m}$, $r_{21}^{max}\approx4.15\,\mu\text{m}$ and $r_{31}^{max}\approx4.6\,\mu\text{m}$, respectively.

\subsubsection{Ellipticity}
Having discussed the perturbations arising from local variations in fiber radii and the core offset, we, here, analyze the modal interactions induced by an elliptical deformation of the fiber cross-section. The governing coupling coefficients are evaluated employing the identical theoretical approach established for the core offset, differing only by the application of the azimuthal harmonic order $N=2$ and its corresponding deformation amplitude $d_{2}$. Assuming a constant ellipticity amplitude of $d_{2} = 0.5\,\mu\text{m}$, the calculated coupling coefficients $\mathbb{K}_{01-21}$ and $\mathbb{K}_{11-31}$ for RCF2 are presented in figure~5(a) and (b), respectively. A similar dependency of DBE efficiency on the azimuthal order of the interacting modes is observed here, confirming our previous findings for modal couplings induced by the core displacement perturbation. Moreover, simulation results demonstrate that higher azimuthal perturbation orders weaken the impact of DBE, meaning that DBE influences modal interactions arising from the core offset perturbation more strongly than those induced by core ellipticity.
\begin{figure}[!t]
\centering\includegraphics[width=9cm]{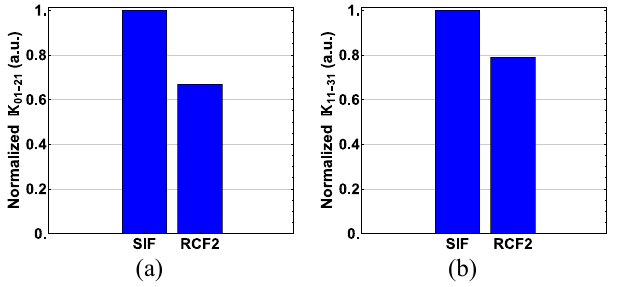}
\caption{Calculated coupling coefficients arising from an elliptical deformation of $d_2 = 0.5\,\mu\text{m}$: (a) $\mathbb{K}_{01-21}$, and (b) $\mathbb{K}_{11-31}$. The numerical aperture of the guiding regions is $\approx 0.29$. As in figure~2, the conventional SIF serves as the reference one, and the coupling coefficients in RCF2 are compared with the corresponding ones in the reference fiber.}
\end{figure}
To explain this behavior, we analyze the spatial overlap of the interacting modes at the ring boundaries. The efficiency of DBE is fundamentally governed by the radial profile overlap between the interacting modes localized at the two interfaces $R_1$ and $R_2$. Figure~6(a) illustrates the radial field distributions of three OAM modes involved in the modal couplings $\mathbb{K}_{01-11}$ (induced by the $N=1$ perturbation) and $\mathbb{K}_{01-21}$ (induced by the $N=2$ perturbation), where $\text{OAM}_{01}$ is common to both interactions. The radial profiles of $\text{OAM}_{01}$ and $\text{OAM}_{11}$ exhibit a stronger spatial overlap at the boundaries compared to those of $\text{OAM}_{01}$ and $\text{OAM}_{21}$. This results in a more pronounced DBE in the $\mathbb{K}_{01-11}$ interaction compared to $\mathbb{K}_{01-21}$. 
\begin{figure}[!t]
\centering\includegraphics[width=13.5cm]{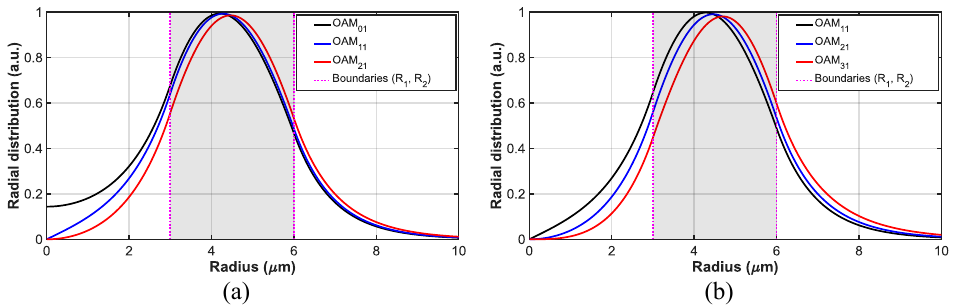}
\caption{Radial electric field distribution of the OAM modes involved in the modal interactions (a) $\mathbb{K}_{01-11}$ and $\mathbb{K}_{01-21}$, and (b) $\mathbb{K}_{11-21}$ and $\mathbb{K}_{11-31}$, in RCF2. $\mathbb{K}_{01-11}$ and $\mathbb{K}_{11-21}$ couplings are induced by the $N=1$ perturbation, while $\mathbb{K}_{01-21}$ and $\mathbb{K}_{11-31}$ couplings are induced by the $N=2$ perturbation.}
\end{figure}
A similar analysis is presented in figure~6(b) for $\mathbb{K}_{11-21}$ and $\mathbb{K}_{11-31}$ couplings, which share the common $\text{OAM}_{11}$ mode. Due to the strong radial profile overlap between $\text{OAM}_{11}$ and $\text{OAM}_{21}$ compared to $\text{OAM}_{11}-\text{OAM}_{31}$ pair, DBE remains more efficient for the $\mathbb{K}_{11-21}$ coupling.
Considering the dependence of the DBE efficiency on the azimuthal orders of both the interacting modes and the structural perturbation, figures~6(a) and 6(b) present only the most representative cases. The results for RCF1 under elliptical deformation are not included because, as demonstrated in the previous section for the core-offset perturbation, the DBE is significantly weaker in RCF1 than in RCF2 due to its geometry being more similar to that of the conventional SIF. As a result, the corresponding coupling coefficients are nearly identical to those of the SIF and do not provide additional physical insight.
\subsubsection{Propagation analysis}
To investigate the impact of DBE on the propagation dynamics and modal purity of OAM modes, we consider a two-mode coupling system governed by the $\mathbb{K}_{01-11}$ interaction between $\text{OAM}_{01}$ and $\text{OAM}_{11}$ modes in both the SIF and RCF2. To isolate the DBE contribution to modal dynamics we assume perfect phase matching ($\Delta\beta = 0$) in both fibers. Physically, this resonant condition can be achieved by precisely tuning the twist rate in helically twisted fibers \cite{asgharzadeh2026optical} or can arise from specific random perturbations.
\begin{figure}[!t]
\centering\includegraphics[width=8cm]{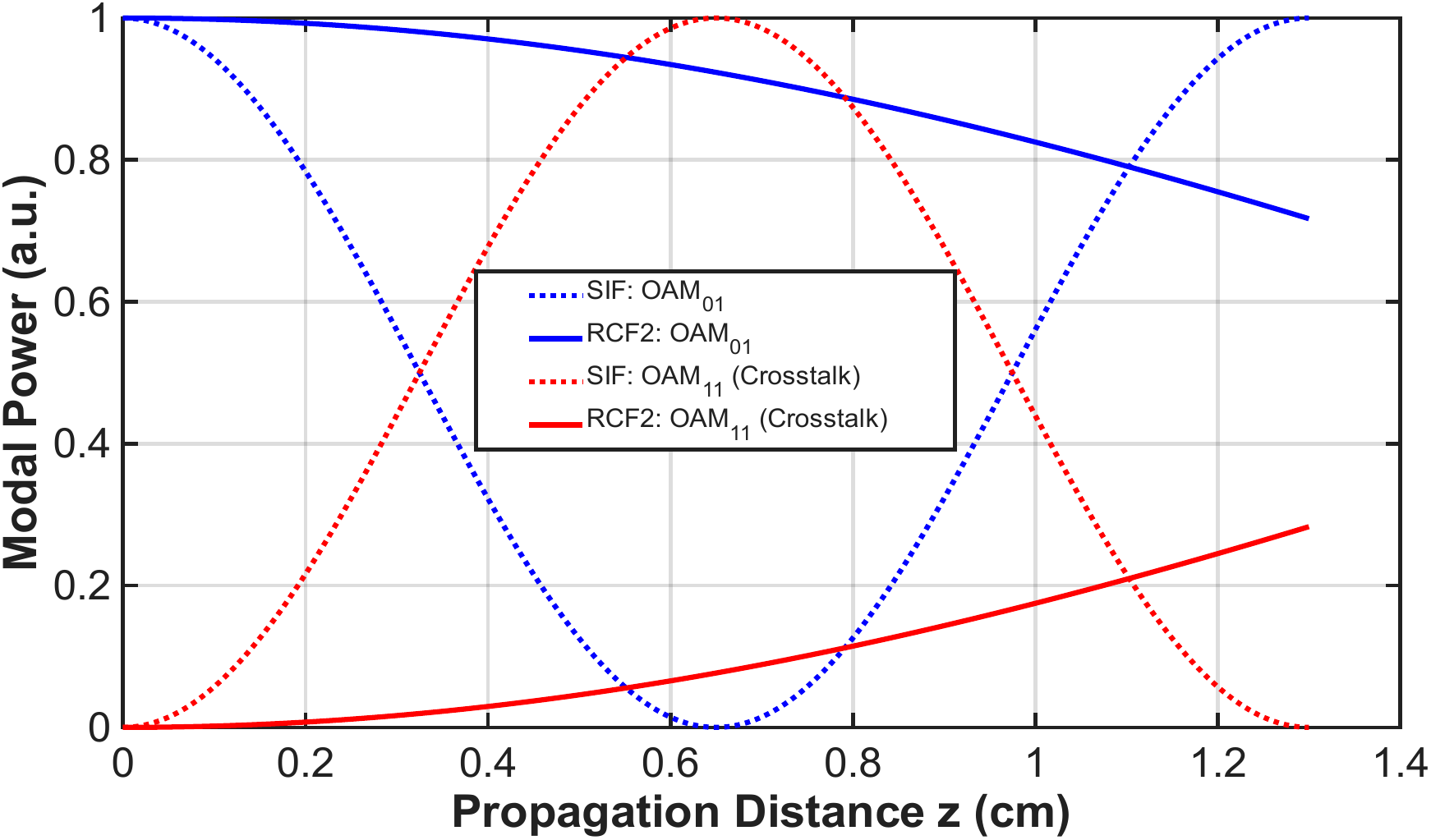}
\caption{Modal cross-talk and topological robustness in SIFs and RCFs. Simulated power evolution under a constant core offset perturbation ($d_{1} = 0.1\,\mu\text{m}$) for SIF and RCF2 cases. The interaction is evaluated under a strict phase-matching condition ($\Delta\beta = 0$). While SIF exhibits complete power transfer to the parasitic $\text{OAM}_{11}$ mode at z = 0.65 cm, RCF2 restricts cross-talk to 8\%, highlighting its superior robustness against geometrical imperfections.}
\end{figure}
We model the geometrical imperfection as a physically realistic uniform core offset with an amplitude of $d_{1} = 0.1\,\mu\text{m}$. Assuming a pure $\text{OAM}_{01}$ excitation at the input (z = 0), figure~7 illustrates the significant mitigation of modal cross-talk in RCF2. In the SIF, complete power transfer to the parasitic $\text{OAM}_{11}$ mode occurs at a propagation distance of z = 0.65 cm. However, at this exact distance, the power transfer in the RCF2 is only 8\%. Furthermore, at z = 1.3 cm, where SIF undergoes a full power return to the fundamental mode, the accumulated cross-talk in RCF2 reaches to approximately 28\%, demonstrating the enhanced topological robustness of RCF geometries against N=1 boundary perturbations.

\section{Conclusion}\label{section4}

In this work, we investigated the influence of DBE on the modal purity and propagation stability of OAM modes in RCFs. To rigorously isolate the physical contribution of the inner boundary, we compared the RCF performance against SIFs possessing identical core radius and numerical apertures. We established a rigorous theoretical framework to evaluate DBE under realistic structural perturbations, specifically addressing radius variations (tapering), core offsets, and elliptical deformations. Our numerical simulations reveal that RCFs fundamentally mitigate asymmetric modal cross-talk. By evaluating the  induced modal interactions, we demonstrated that the net coupling coefficients are significantly suppressed in RCFs, particularly for lower-order modal interactions. As a quantitative demonstration, under strict phase-matching conditions and an identical core-offset perturbation, the maximum parasitic power transfer from the $\text{OAM}_{01}$ to the $\text{OAM}_{11}$ mode is reduced from $100\%$ in the SIF reference to $8\%$ in the RCF. These findings establish that RCFs possess an inherent, geometry-driven topological robustness, making them a highly efficient platform for the propagation of pure OAM beams.

\section*{Acknowledgements}
The authors acknowledge the Research Council of Finland, PREIN Flagship Proggramme (decision No. 320165), and HORIZON EUROPE European Innovation Council, VORTEX4FUSION Project (grant No. 101096317)

\bibliographystyle{iopart-num-long}
\bibliography{iopart-num}

\end{document}